\documentclass[aps,twocolumn,showpacs,preprintnumbers,prl,superscriptaddress,10pt,nofootinbib]{revtex4-1}
\usepackage[utf8]{inputenc}
\usepackage{graphicx}
\usepackage{amsmath,amssymb}
\usepackage{amsfonts}
\usepackage{physics}
\usepackage{xspace}
\usepackage{bm}
\usepackage[normalem]{ulem}
\usepackage{mathrsfs}
\usepackage[dvipsnames]{xcolor}
\usepackage[unicode]{hyperref}
\hypersetup{colorlinks=true, citecolor=MidnightBlue,
	linkcolor=Brown, urlcolor=Blue, linktocpage=true}
\usepackage[all]{hypcap}
\usepackage{multirow}
\usepackage[format=plain,justification=RaggedRight,labelsep=endash,font=small,labelfont=sc]{caption}
\usepackage[export]{adjustbox}

\usepackage{caption}
\usepackage{subcaption}

\definecolor {darkgreen}{rgb}{0.2,0.7,0.2}

\definecolor {dark}{rgb}{0.43,0.5,0.5}

\newcommand{\eq}{\begin{equation}}
\newcommand{\be}{\begin{equation}}
\newcommand{\eeq}{\end{equation}}
\newcommand{\ee}{\end{equation}}

\newcommand{\bea}{\begin{eqnarray}}
\newcommand{\ena}{\end{eqnarray}}

\newcommand{\nb}{\nonumber}

\newcommand{\Stu}{Stueckelberg }

\begin{document}

\title{The well-posedness of the Cauchy problem for self-interacting vector fields}
\author{Enrico Barausse}
\affiliation{SISSA, Via Bonomea 265, 34136 Trieste, Italy and INFN Sezione di Trieste}
 \affiliation{IFPU - Institute for Fundamental Physics of the Universe, Via Beirut 2, 34014 Trieste, Italy}
 \author{Miguel Bezares}
\affiliation{SISSA, Via Bonomea 265, 34136 Trieste, Italy and INFN Sezione di Trieste}
 \affiliation{IFPU - Institute for Fundamental Physics of the Universe, Via Beirut 2, 34014 Trieste, Italy}
 \author{Marco Crisostomi}
 \affiliation{SISSA, Via Bonomea 265, 34136 Trieste, Italy and INFN Sezione di Trieste}
 \affiliation{IFPU - Institute for Fundamental Physics of the Universe, Via Beirut 2, 34014 Trieste, Italy}
\author{Guillermo Lara}
\affiliation{SISSA, Via Bonomea 265, 34136 Trieste, Italy and INFN Sezione di Trieste}
\affiliation{IFPU - Institute for Fundamental Physics of the Universe, Via Beirut 2, 34014 Trieste, Italy}
\affiliation{Max Planck Institute for Gravitational Physics (Albert Einstein Institute), D-14476 Potsdam, Germany}

\begin{abstract}
\noindent
We point out that the initial-value (Cauchy) problem for self-interacting vector fields presents the same well-posedness issues as for first-order derivative self-interacting scalar fields (often referred to as $k$-essence). 
For the latter, suitable 
strategies have been employed in the last few years to
successfully evolve the Cauchy problem at the level
of the infrared theory, without the need for an explicit ultraviolet completion. We argue that the very same techniques
can also be applied to self-interacting vector fields, avoiding a number of issues and ``pathologies'' recently found in the literature.
\end{abstract}
\pacs{}
\date{\today \hspace{0.2truecm}}

\maketitle
\flushbottom

Massive vector fields are ubiquitous in physics, e.g. as mediators of the weak interaction, dark matter candidates and superconductivity.
Recently, considerable interest has grown around possible pathologies that allegedly arise when self-interactions are considered for these fields.
In more detail, in the presence of gravity, 
the action for a real vector field $A_\mu$ with mass $m$ is given by 
\bea
&& S = \int d^4x\sqrt{-g} \left[ \frac{M_{\text{P}}^2}{2} R - \frac14 F_{\mu\nu}F^{\mu\nu} \right. \nb \\
&& \qquad\qquad\qquad\quad \left. -\frac12 m^2 A_\mu A^\mu +\lambda \left(A_\mu A^\mu\right)^2 + \cdots \right] \,, \label{proca}
\ena
where $F_{\mu\nu} \equiv \nabla_\mu A_\nu - \nabla_\nu A_\mu$ and we have introduced the coupling constant $\lambda$ for the lowest order self-interaction (the dots denote higher-order operators).
Ref.~\cite{Clough:2022ygm} has shown
that numerical initial-value (Cauchy) evolutions of such a 
vector field on a black hole background break down, and attributes this feature to the appearance of ghost (or tachyonic) instabilities. 
Soon afterwards, Ref.~\cite{Mou:2022hqb} and \cite{Coates:2022qia} have identified the same problem, although in a simplified set-up.
Similar issues were already pointed out perturbatively on non-vanishing backgrounds in \cite{Barvinskii}.

In this short note, we wish to stress that those pathologies are not surprising when the action is rewritten in the \Stu language, and that
they are actually related to the (breakdown of the) well-posedness of the Cauchy problem in the set-up considered by these works. Indeed, introducing a new scalar field $\phi$, we can restore the U(1) gauge symmetry of the action by performing the transformation
\be
A_\mu \to A_\mu + \frac1m \nabla_\mu \phi \,,
\ee
which renders the longitudinal mode of the massive vector field explicit. We can now choose the ``unitary'' gauge $\phi=0$ and get back the original Lagrangian given by Eq.~(\ref{proca}), or we can choose a different gauge, e.g. the Lorenz gauge $\nabla_\mu A^\mu =0$. The latter is particularly useful when one focuses on the relatively high-energy limit of the theory (where one retains only the highest derivative terms in the action), since it decouples the scalar from the vector field and gives
\bea
&& S = \int d^4x\sqrt{-g} \left[ \frac{M_{\text{P}}^2}{2} R - \frac14 F_{\mu\nu}F^{\mu\nu} \right. \nb \\ 
&& \left. \qquad - \frac12 \nabla_\mu \phi \nabla^\mu \phi + \frac{\lambda}{m^4} \left( \nabla_\mu \phi \nabla^\mu \phi \right)^2 + {\cal O} \left( \frac{\nabla}{m} \right)^3 \right] \,. \label{KX}
\ena
In this form, it is straightforward to realize that the (seemingly innocuous) self-interactions of the original vector field actually hide derivative self-interactions, which modify the principal part of the evolution system. In particular, the \Stu field presents first-order derivative self-interactions, which have been extensively studied in the literature (where they are often referred to as $k$-essence) and shown~\cite{Ripley:2019hxt,Bernard:2019fjb,Figueras:2020dzx,Bezares:2020wkn,terHaar:2020xxb,Bezares:2021dma,Bezares:2021yek}  to cause  problems akin to those encountered in \cite{Clough:2022ygm, Mou:2022hqb, Coates:2022qia}.

These problems arise from the breakdown of strong hyperbolicity (and thus of the well-posedness of the Cauchy problem) for the \Stu field equations.
A system of partial differential equations is strongly hyperbolic if the characteristic matrix of the principal part has real eigenvalues and a complete set of eigenvectors; a
sufficient requirement for this is that
the eigenvalues (i.e. the characteristic speeds) are real and distinct.
Because of the derivative self-interactions in Eq.~(\ref{KX}), the characteristic speeds depend on the scalar field gradients~\cite{Bernard:2019fjb,Bezares:2020wkn,terHaar:2020xxb}, potentially leading to several issues.

If the characteristic speeds cease to be real and distinct along an initial value evolution, the system may become parabolic or elliptic.
An example of this kind~\cite{Stewart:2002vd} 
is provided by the Tricomi equation~\cite{Ripley:2019hxt,Bernard:2019fjb,Figueras:2020dzx}
\begin{eqnarray}
\partial^2_{t}\phi(t,r) + t \, \partial^2_{r}\phi(t,r) = 0~,
\label{trico}
\end{eqnarray}
which is hyperbolic for $t<0$ (as it has characteristic speeds $\pm \, (-t)^{1/2}$) and elliptic for $t>0$.
Furthermore, the characteristic speeds may even
diverge. An example  is given by the Keldysh equation~\cite{Ripley:2019hxt,Bernard:2019fjb,Figueras:2020dzx}
\begin{equation}
t\partial^2_t\phi(t,r) + \partial^2_r\phi(t,r) = 0~,
\label{keld}
\end{equation}
which has  characteristic speeds $\pm \, (-t)^{-1/2}$
diverging as $t\to0^-$.
Finally, it has been shown that 
derivative self-interactions can lead to the formation of shocks/microshocks even when starting from smooth initial data, potentially leading to non-unique solutions and therefore to an ill-posed Cauchy problem~\cite{Reall:2014sla,Tanahashi:2017kgn}.

Solutions to these issues, however, have been put forward in recent years. Ref.~\cite{Bezares:2020wkn} has shown
that  the Tricomi-type evolution system
arising from the action of Eq.~\eqref{KX}
can 
be avoided altogether if the coefficients of the derivative self-interactions
satisfy suitable conditions.
For example, accounting for a cubic term
 $(\nabla_\mu \phi \nabla^\mu \phi)^3$
 is sufficient to avoid the loss of hyperbolicity.

Moreover, Ref.~\cite{Bezares:2020wkn} and \cite{terHaar:2020xxb} have shown 
that for theories with this cubic term, 
a Keldysh-type breakdown  
of the Cauchy evolution
typically occurs only  during black hole collapse,
for realistic initial data.
However, 
the diverging
characteristic speeds that define the Keldysh behavior
are not pathological {\it per se}, but are simply due to 
a poor choice of gauge.
Indeed, Ref.~\cite{Bezares:2021dma} has found a gauge  (with non-vanishing shift) that maintains the characteristic speeds finite 
in stellar oscillations, during gravitational collapse and in binary neutron star mergers.
Finally, the problem of shock formation can be avoided by writing the evolution equations
as a hyperbolic 
conservation 
system,
and by solving the 
latter using high-resolution shock capturing
techniques~\cite{Bezares:2020wkn, terHaar:2020xxb, Bezares:2021dma}.

The discriminant choice to perform numerical evolutions in self-interacting vector theories is therefore a careful selection of the coupling constants, so as to satisfy the conditions 
avoiding Tricomi type breakdowns of well-posedness \cite{Bezares:2020wkn}. If such conditions are not satisfied (as for the cases studied in \cite{Clough:2022ygm, Mou:2022hqb, Coates:2022qia}) an alternative possibility is to employ 
a ``fixing-equation''~\cite{Cayuso:2017iqc} approach inspired  by  the M\"{u}ller–Israel–Stewart formulation of viscous relativistic hydrodynamics~\cite{Muller:1967zza,Israel:1976tn,1976PhLA...58..213I}. These approaches have been successfully applied in \cite{Cayuso:2017iqc,Allwright:2018rut,Cayuso:2020lca, Bezares:2021yek, Gerhardinger:2022bcw,Franchini:2022ukz,Lara:2021piy} to ameliorate the stability of the fully numerical evolutions in theories with either changes of character of the Cauchy problem, higher-order derivatives, or derivative self-interactions.
This method consists of modifying the 
field equations by introducing extra fields and 
``fixing equations'' (i.e. drivers) for them.
The fixing equations are devised
such that 
on long timescales the evolution  approximately matches that of the original effective field theory.
Another possibility is to rely on the ultraviolet (UV) completion of the theory (when that is known) to continue the evolution past the Tricomi or Keldysh breakdown \cite{Lara:2021piy}. 
Unfortunately, for the most interesting cases
(e.g. derivative self-interactions yielding screening mechanisms) a UV completion is not generally known.

\textit{Note added:} When this note was being completed, another paper \cite{Aoki:2022woy} appeared on the arXiv which also introduces the \Stu formulation for self-interacting vector fields.

\textit{Acknowledgments:} 
We thank C.~Palenzuela and  R.~Aguilera-Miret for insightful conversations on numerical relativity and the Cauchy problem. We acknowledge support from the European Union's H2020 ERC Consolidator Grant ``GRavity from Astrophysical to Microscopic Scales'' (Grant No.  GRAMS-815673) and the EU Horizon 2020 Research and Innovation Programme under the Marie Sklodowska-Curie Grant Agreement No. 101007855.

\bibliographystyle{utphys}
\bibliography{master}
\end{document}